\documentclass{elsarticle}

\usepackage{amsmath}
\usepackage{amsfonts}
\usepackage{amssymb,latexsym}
\usepackage{amsthm}
\usepackage{graphicx}
\usepackage{ulem}
\usepackage{color}
\usepackage{newlfont} 
\usepackage{subfigure}
\usepackage{verbatim}
\usepackage{rotating}
\usepackage{multirow}
\usepackage{units}
\usepackage{bm}
\usepackage[english]{babel}
\usepackage[utf8]{inputenc}
\usepackage{algorithm}
\usepackage[noend]{algpseudocode} 
\usepackage{booktabs}
\usepackage{caption}
\usepackage{hyperref}
\usepackage{subfigure}
\usepackage[export]{adjustbox}

\newtheorem{theorem}{Theorem}[section]

\newtheorem{Remark}{Remark}[section]

\begin{document}

\begin{frontmatter}
\title{ORTHOCUB: \\
integral and  differential cubature rules\\ by orthogonal moments}
\author[address-PD]{Laura Rinaldi}
\ead{laura.rinaldi@math.unipd.it}

\author[address-PD]{Alvise Sommariva}
\ead{alvise@math.unipd.it}

\author[address-PD]{Marco Vianello}
\ead{marcov@math.unipd.it}

\address[address-PD]{University of Padova, Italy}

\begin{abstract}
We discuss a numerical package, named ORTHOCUB, for the computation of linear functionals of both integral and differential type on multivariate polynomial spaces. The weighted sums corresponding to such integral and differential cubatures are implemented via orthogonal polynomial moments and auxiliary near-minimal algebraic cubature in a bounding box, with no conditioning issue since no matrix inversion or factorization is needed. The whole computational process indeed reduces to moment computation and dense matrix-vector products of relatively small  size. 
The Matlab and Python codes are freely available, to be used as building blocks for integral and differential problems. 

\vspace{0.5cm} {\it{Keywords: 65D25, 65D30, 65D32}}
\end{abstract}
	

\end{frontmatter}
 
\section{Introduction}
In the recent paper \cite{RSV25}, we have shown that given an integration functional $\mathcal{L}(f)=\int_\Omega{f(P)\,d\nu}$, the availability of an orthonormal polynomial basis $\{\phi_j\}$ in a bounding set $B\supseteq \Omega$ together with an algebraic cubature formula for the orthogonality measure (say $\mu$), and of the corresponding orthogonal moments 
$\mathbf{m}=\{\mathcal{L}(\phi_j)\}$, allows to represent cheaply the functional as a weighted sum of the function values sampled at the nodes of the underlying cubature formula. The weight computation can be made avoiding at all matrix inversions (linear systems) and factorizations, eliminating at the root any conditioning issue, and accelerating the computation with respect to other methods seeking for positive weights. 

Moreover, though the weights are not all positive, we have shown that if $\mathbf{w}$ are the weights and $\mathbf{m}$ the moments then $\|\mathbf{w}\|_1\leq \sqrt{\mu(B)}\|\mathbf{m}\|_2$, 
which allowed to study the stability of the resulting discretized functional by estimating $\|\mathbf{m}\|_2$ via the classical Bessel and Jensen inequalities. The method can be applied for example to the evaluation of integrals of polynomials on complex shaped elements for FEM/VEM in 2D and 3D, or for the compression of large scale QMC computations on difficult 3D geometries. 
  
In the present paper we specialize the implementation to the case of a bounding box $B$, but we also extend to {\em general linear functionals}, thus allowing to treat for example differentiation functionals besides integration and summation 
functionals, namely not only
\begin{equation} \label{int-fun}
\mathcal{L}(f)=\int_\Omega{f(P)}\,d\nu\;,\;\;\Omega\subseteq B\subset \mathbb{R}^d
\end{equation}
where $\nu$ is a continuous or discrete measure, but also 
\begin{equation} \label{diff-fun}
\mathcal{L}(f)=\partial^\alpha f(P)\;,\;\;P=(x_1,\dots,x_d)\in B\;,
\end{equation}
corresponding to (the pointwise evaluation of)  a partial differential operator $\partial^\alpha=\partial^{\alpha_1}_{x_1}\,\dots 
\,\partial^{\alpha_d}_{x_d}$. 

In Section 2 we pose the theoretical base of the method, whereas in Section 3 we discuss its implementation in the case of 2D and 3D boxes $B$, via near-minimal algebraic cubature formulas for the underlying orthogonality measure. The corresponding Matlab and Python codes, collected in a package named ORTHOCUB, are freely available. Finally, in Section 4 we describe some demos concerning both, integration functionals on complex shaped elements in view of FEM applications, and differential functionals in view of numerical differentiation as well as discretization of PDEs by differential cubature.

\section{Weighted sum functionals by orthogonal moments}

Following \cite{RSV25}, we can state and prove the following general representation result for linear functionals on polynomial spaces.

\begin{theorem}
 Let $B$ be (the closure of) an open set in $\mathbb{R}^d$ and $d\mu=\sigma(P)\,dP$ an absolutely continuous measure with respect to the Lebesgue measure on $B$.  Moreover, let $\{\phi_j\}_{1\leq j\leq N}$ be a $\mu$-orthonormal basis for $\mathbb{P}_n$ (the polynomials of total-degree not exceeding $n$), where $N=dim(\mathbb{P}_n)={n+d \choose d}$, 
and let $(X,\mathbf{u})=\{(P_i,u_i),\,1\leq i\leq M\}$, $M\geq N$, be the nodes $X\subset B$ and positive weights of a quadrature formula with Algebraic Degree of Exactness $ADE=2n$ for $\mu$, that is 
\begin{equation} \label{exact2n}
\int_B{\phi(P)\,d\mu}=\sum_{i=1}^M{u_i\,\phi(P_i)}\;,\;\;\forall \phi\in \mathbb{P}_{2n}\;.
\end{equation}

Let $\mathcal{L}: S\to \mathbb{R}$ a linear functional, where $S$ is a subspace of real-valued functions on $B$ containing the polynomials. 
Consider the Vandermonde-like matrix and the diagonal weight matrix 
\begin{equation} \label{vand}
{\color{black}V_\phi=}(v_{ij})=(\phi_j(P_i))\in \mathbb{R}^{M\times N}\;,\;\;diag(\mathbf{u})\in \mathbb{R}^{M\times M}\;,
\end{equation}
and denote by $\mathbf{m}=\{m_j\}\in \mathbb{R}^N$ the vector of moments
\begin{equation} \label{mom}
m_j=\mathcal{L}(\phi_j)\;,\;\;1\leq j\leq N\;.
\end{equation}

Then, the discrete signed measure $(X,\mathbf{w})=\{(P_i,w_i),\,1\leq i\leq M\}$ 
supported on $X\subset B$, with weights $\mathbf{w}=\{w_i\}\in \mathbb{R}^M$ computed as 
\begin{equation} \label{w}
\mathbf{w}=diag(\mathbf{u})V_\phi\mathbf{m}\;,
\end{equation}
gives an exact ``cubature formula'' in $\mathbb{P}_n$ for the functional $\mathcal{L}$ 
\begin{equation} \label{exact}
\mathcal{L}_n(f)=
\sum_{i=1}^M{w_i\,f(P_i)}=\mathcal{L}(f)\;,\;\;\forall f\in \mathbb{P}_{n}\;,
\end{equation} 
with the following bound for the weights  
\begin{equation} \label{bound}
\;\;\|\mathbf{w}\|_1\leq \sqrt{\mu(B)}\,\|\mathbf{m}\|_2\;\;.
\end{equation}   
\end{theorem}

\vskip0.5cm
\noindent
{\em Proof.} We follow the same reasoning of \cite{RSV25}, applied here to a general linear functional. Let us put for simplicity $V=V_\phi$ and $D=diag(\mathbf{u}) $. By linearity of $\mathcal{L}$ and $\mathcal{L}_n$, condition (\ref{exact}) is satified if and only if $\mathcal{L}_n(\phi_j)=\mathcal{L}(\phi_j)$ for every $j$, which is in turn equivalent to the fact that the weights solve the (underdetermined) linear system $V^ t\mathbf{w}=\mathbf{m}$. The latter holds true since by (\ref{w}), orthonormality of the basis and exactness degree $ADE=2n$ of the cubature formula $(X,\mathbf{u})$, we have that the matrix $A=D^{1/2}V$ is orthogonal and thus $V^t\mathbf{w}=V^tDV\mathbf{m}=A^ tA\mathbf{m}=\mathbf{m}$. 

On the other hand, bound (\ref{bound}) holds by the chain of inequalities 
$$
\|\mathbf{w}\|_1=\sum_{i=1}^M{|w_i|}=\sum_{i=1}^M{(|w_i|/\sqrt{u_i}\;)\sqrt{u_i}}\leq \|\{|w_i|/\sqrt{u_i}\,\}\|_2\,\|\{\sqrt{u_i}\,\}\|_2
$$
$$
=\|D^{-1/2}\mathbf{w}\|_2\,\sqrt{\sum_{i=1}^M{u_i}}=\|A\mathbf{m}\|_2\,\sqrt{\mu(B)}=\|\mathbf{m}\|_2\,\sqrt{\mu(B)}\;.\hspace{1cm} \square
$$

\section{Implementation}

In this section we focalize on the case when $B$ is a {\em Cartesian box}. After choosing an orthonormal polynomial basis $\{\psi_j\}$ in the reference cube $[-1,1]^d$, in view of the affine change of variables from $[-1,1]^d$ to the box $B$, namely 
$$
Q=(t_1,\dots,t_d)
\mapsto P=(x_1,\dots,x_d)
$$
\begin{equation} \label{P&Q}
P=\Lambda Q+C\;,\;\;\Lambda=diag(\lambda_1,\dots,\lambda_d)\;,\;C=(c_1,\dots,c_d)
\end{equation}
(where the scaling values $\{\lambda_k\}$ are the half-lengths of the box sides and $C$ is the box center), we can work with the orthonormal basis 
\begin{equation}\label{10}
\phi_j(P)=\sqrt{det(\Lambda^{-1})}\,\psi_j(\Lambda^{-1}(P-C))
\end{equation}
in $B$. In such a way, given a cubature formula in $[-1,1]^d$ with $ADE=2n$ for the orthogonality measure, say $(\{Q_i\},\mathbf{z)}$, we get a formula $(X,\mathbf{u})$ in $B$ with 
\begin{equation}
X=\{P_i\}=\{\Lambda Q_i+C\}\;,\;\;\mathbf{u}=det(\Lambda)\, \mathbf{z}\;, 
\end{equation}
and finally the ORTHOCUB weights associated  to the nodes $X$
{\color{black}
\begin{equation}
\mathbf{w}=det(\Lambda)\,diag(\mathbf{z})\,V_\phi\mathbf{m}=det(\Lambda)\,diag(\mathbf{z})\,\sqrt{det(\Lambda^{-1})}\,V_\psi\mathbf{m}=\sqrt{det(\Lambda)} diag(\mathbf{z})\,V_\psi\mathbf{m}\;.
\end{equation}
}

We stress that the matrix 
\begin{equation} \label{V-once}
diag(\mathbf{z})V_\psi=(z_i\psi_j(Q_i))\;,\;1\leq i\leq M\;,\;1\leq j\leq N\;,
\end{equation}
can be computed {\em once and for all} for a given $n$ in the reference box $[-1,1]^d$, and the overall ORTHOCUB weights computation, in view of (\ref{w}) and the fact that
\begin{equation}
\mathbf{m}=\{\mathcal{L}(\phi_j)\}=\{\mathcal{L}\left(\sqrt{det(\Lambda^{-1})} \psi_j(\Lambda^{-1}(\cdot-C))\right)\}=\sqrt{det(\Lambda^{-1})}\,\{\mathcal{L}\left(\psi_j(\Lambda^{-1}(\cdot-C))\right)\}\;,
\end{equation}
boils down to 
{\color{black}
\begin{equation} \label{weights2}
\mathbf{w}=\sqrt{det(\Lambda)}\,diag(\mathbf{z})\,V_\psi\mathbf{m}=diag(\mathbf{z})\,V_\psi\{\mathcal{L}\left(\psi_j(\Lambda^{-1}(\cdot-C))\right)\}
\end{equation}
}
that is nothing but a {\em single matrix-vector product} with a {\em fixed matrix}. As already observed, this is exactly the strength of the ORTHOCUB method for the representation of linear functionals as weighted sums of function values, since {\em no linear system} has to be solved and {\em no matrix factorization} computed, as instead required by other weighting methods in the integral or differential frameworks, thus having at the same time {\em no conditioning problem} in the weights computation. 

\subsection{Near-minimal rules for the product Chebyshev measure}

We have specialized the present implementation of ORTHOCUB to 2D and 3D boxes, using a classical family of product orthonormal polynomials on the reference box $[-1,1]^d$, 
$d=2,3$. Given the normalized Chebyshev polynomials of the first kind 
\begin{equation}
p_s(t)=\sqrt{\frac{2-\delta_{s,o}}{\pi}}\;T_s(t)\;,\;T_s(t)=\cos{(s\,\arccos{(t)})}\;,\;s\geq 0\;,
\end{equation}
\begin{itemize}
\item
in the square $[-1,1]^2$ 
$$
\psi_j(t_1,t_2)=p_h(t_1)p_k(t_2)\;,\;\;0\leq h+k\leq n\;,
$$
\begin{equation} \label{cheb2D}
d\mu=\left((1-t_1^2)(1-t_2^2)\right)^{-1/2}dt_1\,dt_2\;,
\end{equation}
where $j=1,\dots,N=(n+1)(n+2)/2$ correspond to a suitable ordering of the $(h,k)$ bi-indexes (we have adopted the graded lexicographical ordering)

\item similarly, in the cube $[-1,1]^3$ 
$$
\psi_j(t_1,t_2,t_3)=p_h(t_1)p_k(t_2)p_\ell\mathcal{L}(t_3)\;,\;\;0\leq h+k+\ell\leq n\;,
$$
\begin{equation} \label{cheb3D}
\;d\mu=\left((1-t_1^2)(1-t_2^2)(1-t_3^2)\right)^{-1/2}dt_1\,dt_2\,dt_3\;,
\end{equation}
where $j=1,\dots,N=(n+1)(n+2)(n+3)/6$ correspond to the same ordering of the $(h,k,\ell)$ tri-indexes. 
\end{itemize}

Clearly, for the purpose of computational efficiency it is convenient to use cubature formulas with $ADE=2n$ for such measures that are {\em minimal}, i.e. with cardinality $M=N$; cf. \cite{M75,X25}. Indeed, one could use tensorial Gaussian rules in $[-1,1]^d$ for any Jacobi measure and any $d$, which however have cardinality $(n+1)^d$. Unfortunately, minimal rules are known theoretically in very few cases, and have been computed numerically for a quite limited range of measures and degrees, cf. \cite{C03,X25}. In alternative, one can resort in many cases to {\em near-minimal formulas}, whose cardinality is close or not far from the minimal one, growing at worst asymptotically as $n^d/d!$ in $n$ for fixed $d$ versus the $n^d$ asymptotics of tensorial rules. Correspondingly to the choices above, in the implementation of ORTHOCUB we have adopted 

\begin{itemize}

\item the near-minimal rules for the product Chebyshev measure of the first kind in $[-1,1]^2$, supported at the MPX (Morrow-Patterson-Xu) points, with cardinality $(n+1)(n+3)/2$ for $n$ odd and $(n+2)^2/2$ for $n$ even (cf. \cite{CDMMV06,X25} and and \cite{S25}; no limitation on the degree);
    
\item the near-minimal rules for the product Chebyshev measure of the first kind in $[-1,1]^3$, 
with cardinality close to $(n+2)^3/4$ (cf. \cite{DMVX09} and \cite{S25}; no limitation on the degree).
\end{itemize}

\subsection{ORTHOCUB and Hyperinterpolation}
It is worth observing that if 
\begin{equation} \label{dirac}
\mathcal{L}(f)=f(P)
\end{equation}
is the evaluation functional at a fixed $P$, then $\mathbf{m}=\mathbf{m}(P)=\{\phi_j(P)\}$ and 
\begin{equation} \label{hyper}
\mathcal{L}_n(f)=
\sum_{i=1}^M{w_i(P)\,f(P_i)}\;,\;\;\mathbf{w}(P)=diag(\mathbf{u})\,V_\phi\{\phi_j(P)\}\;,
\end{equation}
is nothing but Sloan's hyperinterpolant \cite{Slo95} evaluated at $P$, namely
\begin{equation} \label{sloan-h}
\mathcal{H}_n f (P)=\sum_{j=1}^N{\langle f,\phi_j\rangle_{\ell^2_{\mathbf{u}}(X)}\,\phi_j(P)}=\sum_{i=1}^M{\left(u_i\sum_{j=1}^N{\phi_j(P_i)\phi_j(P)}\right)\,f(P_i)}\;,
\end{equation}
i.e. the orthogonal projection on $\mathbb{P}_n$ with respect to the discrete measure corresponding to 
the algebraic cubature formula $(X,\mathbf{u})$ with $ADE=2n$ for the absolutely continuous measure $d\mu=\sigma(P)dP$ on $B$.
Notice that for $M>N=dim(\mathbb{P}_n)$ the polynomials $w_i(P)$ are linearly dependent, hence they do not form a basis but a set of generators of $\mathbb{P}_n$, whereas for $M=N$ (minimal cubature formulas) they form a basis of $\mathbb{P}_n$ and hyperinterpolation becomes standard interpolation. From this point of view, 
ORTHOCUB weighted sum representation of a general functional $\mathcal{L}$ as in (\ref{exact}), can be simply seen as the application of the functional to the hyperinterpolant
\begin{equation} \label{orthocub-h}
\mathcal{L}_n(f)=\mathcal{L}(\mathcal{H}_n f)\;, 
\end{equation}
so that in cases (\ref{int-fun}) and (\ref{diff-fun}), ORTHOCUB is just the integration or differentiation of $\mathcal{H}_n f$. The whole ORTHOCUB software package can then be interpreted as an implementation of hyperinterpolation operators in 2D and 3D boxes, together with the composition of linear functionals with such operators.

It is also worth recalling that the ``Lebesgue constant'' of the hyperinterpolation operator $\mathcal{H}_n:C(B)\to \mathbb{P}_n$, $\mathcal{H}_n f=\sum_{i=1}^M{w_i(\cdot)\,f(P_i)}$, i.e. its uniform norm,  is
\begin{equation} \label{hypernorm}
\|\mathcal{H}_n\|=\sup_{f\neq 0}{\frac{\|\mathcal{H}_n f\|_B}{\|f\|_B}}=\max_{P\in B}{\sum_{i=1}^M{|w_i(P)|}}\;, 
\end{equation}
where $\|f\|_B$ denotes the sup-norm on $B$, cf. e.g. \cite{BCKSV24}.
The growth of this quantity with $n$ depends on the orthogonality measure, and plays a key role for both convergence and stability of the hyperinterpolants. For example, it is known that in the case of the Lebesgue measure 
$\|{\mathcal{H}}_n\|=\mathcal{O}(n^d)$, 
whereas with the product Chebyshev measure of the first kind 
\begin{equation} \label{chebleb}
\|\mathcal{H}_n\|=\mathcal{O}(\log^d(n))\;,
\end{equation}
which is the optimal order of growth for a polynomial projection operator; cf. \cite{DMSV14,SzVert09,WWW14}.

\section{Applications and demos}

\subsection{Integral cubature}
A natural application of ORTHOCUB is the computation of integration (summation) functionals like (\ref{int-fun}), where $\nu$ is either a continuous or a discrete measure on 
$\Omega\subseteq B$. This approach was adopted in \cite{RSV25,SV25} for cheap and stable integration of polynomials on 2D spline elements and 3D polyhedral elements, as well as for the compression of QMC integration on complex-shaped 3D elements. A relevant framework for these computations is that of high-order polytopal FEM/VEM methods, where only the polynomial moments change on each mesh element, while the matrix  $diag(\mathbf{z})V_\psi$ in (\ref{V-once}) can be computed once 
and for all for a whole finite-element mesh. The computational effort then is then essentially given by evaluting of the orthogonal moments
\begin{equation}
\mathcal{L}\left( \psi_j(\Lambda^{-1}(\cdot-C)) \right)= \int_\Omega{\psi_j(\Lambda^{-1}(P-C))\,d\nu}\;,\;1\leq j\leq N\;,
\end{equation}
on the different elements $\Omega$.

As shown in \cite{RSV25} by inequality (\ref{bound}), if $\nu$ is an absolutely continuous measure with respect to the Lebesgue measure on $\Omega$ with density $\omega$, i.e. $d\nu=\omega(P)\,dP$, then $\|\mathbf{w}\|_1$ is bounded in $n$ as soon as $\omega^2/\sigma\in L^1(\Omega)$, which entails that the ORTHOCUB cubature rule is stable (we recall that $\sigma$ is the density of the Chebyshev measure of the first kind on the bounding box $B\supseteq \Omega$). 

On the other hand, if $\nu$ is a general (even discrete) measure on $\Omega$, in \cite{RSV25} it has been proved by Jensen inequality that $\|\mathbf{w}\|_1=\mathcal{O}\left(\max_{P\in B}\sqrt{\sum_{j=1}^N{\phi_j^2(P)}}\right)=\mathcal{O}(n^d)$. This bound can be however refined since, as shown above in \S 3.2, ORTHOCUB corresponds in practice to integrating the hyperinterpolant $\mathcal{H}_n f$, namely for $f\in C(B)$
$$\mathcal{L}_n(f)=\sum_{i=1}^N{w_i\,f(P_i)}=\mathcal{L}(\mathcal{H}_n f)=\int_\Omega{\mathcal{H}_n f(P)\,d\nu}$$ 
and thus $$|\mathcal{L}_n(f)|\leq \nu(\Omega)\|\mathcal{H}_n\|\,\|f\|_{B}\;,$$
which gives by (\ref{hypernorm})
\begin{equation} \label{log}
\|\mathcal{L}_n\|=\|\mathbf{w}\|_1=\mathcal{O}(\log^d(n))\;.
\end{equation}
Our numerical tests have shown, however, that the actual stability of ORTHOCUB for discrete measures is much better than what can be predicted by (\ref{chebleb}) 
(see e.g. Table 2 below).

It is also worth noticing, incidentally, that $\nu$ needs not to be a ``volume'' measure, but could also be a continuous 
or discrete measure on a lower-dimensional manifold. We reserve to future work a full analysis and implementation of ORTHOCUB for the efficient evaluation of surface integrals, possibly on piecewise smooth complicated surfaces, with computational advantages for example with respect to the compression methods developed in \cite{ESV24,ESV24-2}.

\begin{figure}[h]
	\centering
		\includegraphics[scale=0.53,clip,valign=c]{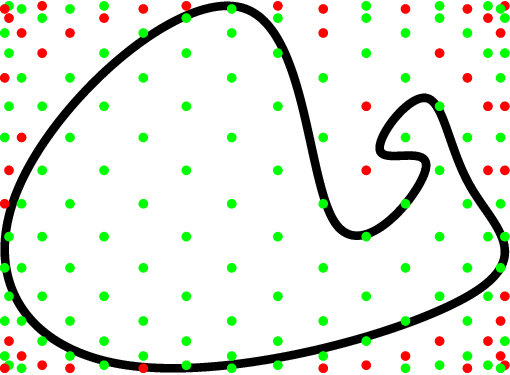}
        \includegraphics[scale=0.36,clip,valign=c]{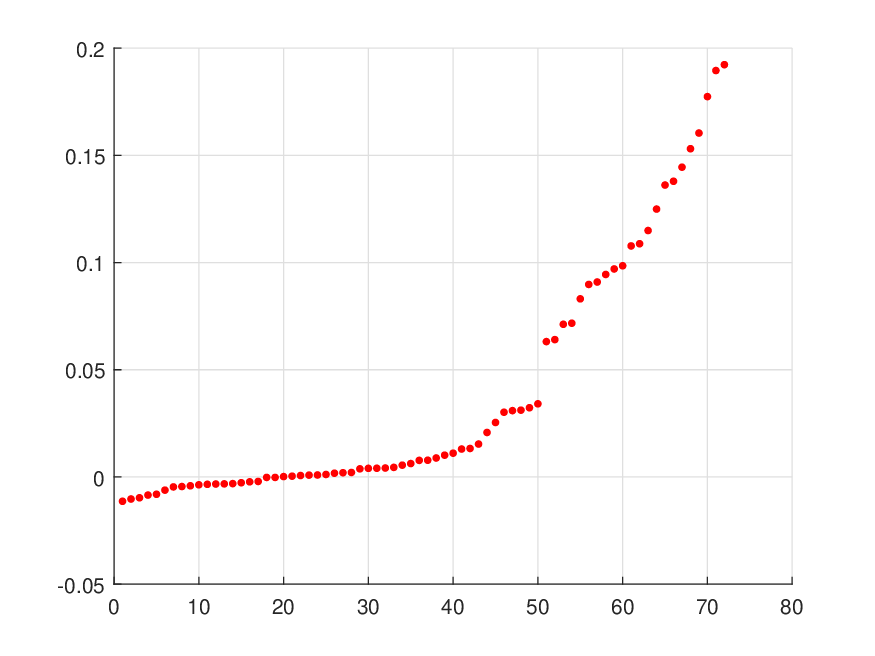}
        \includegraphics[scale=0.36,clip,valign=c]{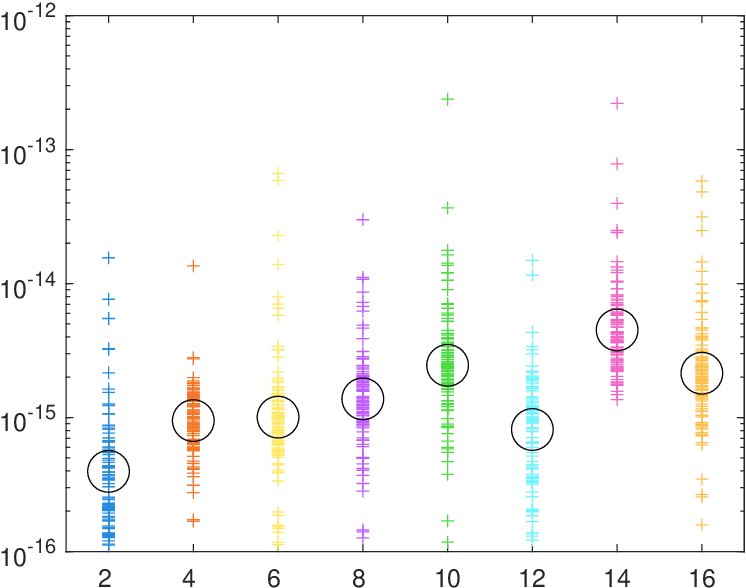}
\caption{ORTHOCUB integration on a curved element with spline boundary (green dots: nodes
with positive weights; red dots: nodes with negative weights): sign and size distribution of the weights for ADE $n=10$, relative integration errors (crosses) for 100 trials of random polynomials $p_n(x,y)=(c_0+c_1 x+c_2 y)^n$, $n=2,4,\dots,16$, with their geometric mean (circles).}
\label{spline}
\end{figure}

We make now a couple of illustrative examples. In the first we apply ORTHOCUB to the Lebesgue integration functional in 2D
\begin{equation} 
\mathcal{L}(f)=\int_\Omega{f(x,y)\,dxdy}
\end{equation}
where $\Omega$ is a curvilinear element with cubic spline boundary. The orthogonal moments in (\ref{weights2}) become by Green theorem 
\begin{equation} 
\mathcal{L}\left(\psi_j(\Lambda^{-1}(\cdot-C))\right)=\lambda_1^{-1}\oint_{\partial \Omega}{\Psi_j(\Lambda^{-1}((x,y)-C))\,dy}\;,\;1\leq j\leq N\;,
\end{equation}
where the polynomial $\Psi_j$ is a primitive of $\psi_j$ with respect to the first variable, i.e. $\partial_{t_1}\Psi_j(t_1,t_2)=\psi_j(t_1,t_2)$, which is known analitycally by explicit formulas for indefinite integrals of univariate Chebyshev polynomials, cf. e.g. \cite{MH03}. Such moments can then be computed efficiently by piecewise Gauss-Legendre quadrature on the cubic pieces of the spline boundary. 
In Figure \ref{spline} we show sign and size distribution of the ORTHOCUB weights for ADE $n=10$. As discussed in \S 3.1, the support of the rule are the MPX points of the box. Moreover, we also display the geometric mean of the relative errors in the integration of a family of $n$-th degree polynomials with random coefficients in $(0,1)$. Such means stay around $10^{-15}$. 

The second example, taken from \cite{RSV25}, concerns the application of ORTHOCUB to a discrete integration functional in 3D (a large scale summation problem), namely cheap evaluation of the QMC integration functional 
\begin{equation} \label{QMC}
S_L(f)=\frac{vol(B)}{L}\,\sum_{k=1}^K{f(\alpha_k,\beta_k,\gamma_k)}\approx \frac{vol(\Omega)}{K}\,\sum_{k=1}^K{f(\alpha_k,\beta_k,\gamma_k)}\approx \int_\Omega{f(x,y,z)}\,dxdydz\;,
\end{equation}
on a complex shaped domain $\Omega$ given by the union of 5 intersecting balls, where $\{(\alpha_k,\beta_k,\gamma_k)\}$ are the fraction falling in $\Omega$ of a large number of Halton points, say $L$, in the bounding box. In (\ref{QMC}) we have used the natural QMC approximation $vol(\Omega)\approx vol(B)\,K/L$. Starting with $L=10^5$ points in the bounding box, we get $K>37000$ points in the ball union.
In this case the orthogonal moments in (\ref{weights2}) are simply 
\begin{equation} 
S_L\left(\psi_j(\Lambda^{-1}(\cdot-C))\right)=\frac{{vol}(B)}{L}\,\sum_{k=1}^K{\psi_j(\Lambda^{-1}((\alpha_k,\beta_k,\gamma_k)-C))}\;,\;1\leq j\leq N\;.
\end{equation}

\begin{figure}[h]
	\centering
		\includegraphics[scale=0.70,clip,valign=c]{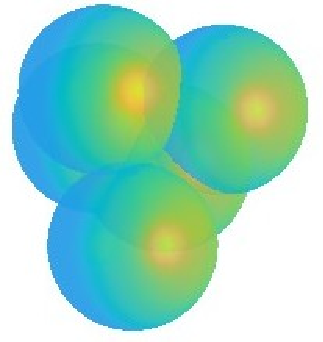}
        \hspace{0.925cm}
        \includegraphics[scale=0.36,clip,valign=c]{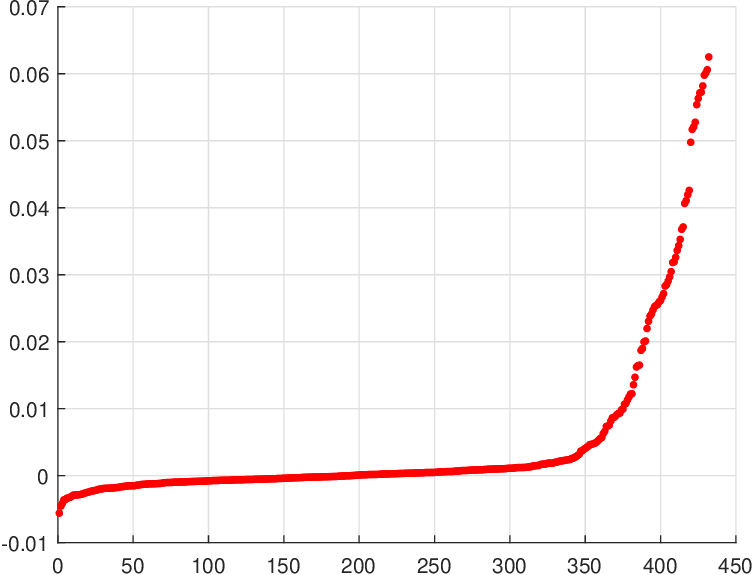}
        \hspace{0.325cm}
        \includegraphics[scale=0.36,clip,valign=c]{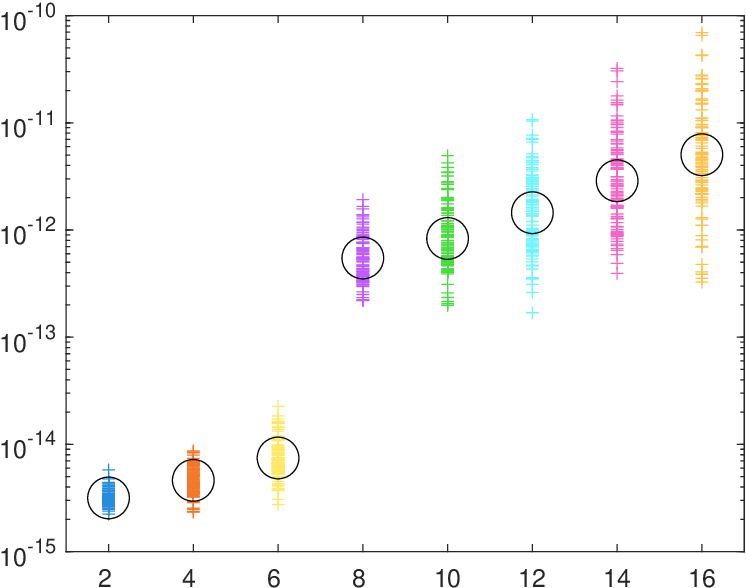}
\caption{ORTHOCUB compression of QMC integration on the union of 5 balls: sign and size distribution of the weights for ADE $n=10$, relative integration errors (crosses) for 100 trials of random polynomials $p_n(x,y,z)=(c_0+c_1 x+c_2 y+c_3z)^n$, $n=2,4,\dots,16$, with their geometric mean (circles).}
\label{balls}
\end{figure}

Again, in Figure \ref{balls} we show sign and size distribution of the ORTHOCUB weights for ADE $n=10$, along with the geometric mean of the relative errors with respect to the QMC formula (\ref{QMC}) for a family of $n$-th degree polynomials with random coefficients in $(0,1)$. We see that such errors range from an order of $10^{-14}$ to an order of $10^{-12}$ at the highest degrees. 

In Table 1, we report the cardinalities of ORTHOCUB for the spline element and for the ball union. We notice that such cardinalities are significantly lower than those in \cite{RSV25}, by roughly a factor $1/2$ in 2D and $1/4$ in 3D, because there the support of the orthogonal cubature is that of a tensorial Gauss-Chebyshev rule with $(n+1)^d$ points. Observe that for QMC integration these cardinalities have to be compared with the more than 37000 points of the original QMC formula. It is worth stressing once more that such a compression, which is particularly impressive in the QMC case, has been obtained without solving any linear system or computing any matrix factorization, differently from other cubature compression techniques, like e.g. those adopted in \cite{ESV24,LvPBD24,SV21} (see also the references therein). In the case of spline boundaries, the sampling cardinality is also much lower than that required by other popular methods in the FEM/VEM literature, like e.g. \cite{SV09}. 

Moreover, in Table \ref{ratios} we display the stability ratios $\|\mathbf{w}\|_1/|\sum_i{w_i}|\approx \|\mathbf{w}\|_1/vol(\Omega)$, which turn out to stay close to the optimal value 1, ensuring a good stability of the corresponding ORTHOCUB rules. 
Finally, in Table \ref{times} we display the CPU times for the construction of the ORTHOCUB  integral cubature weights.

\begin{table} 
\centering
\begin{tabular}{| c || c | c | c || c | c | c ||}
		\hline 
		ADE $n$ & spl bnd & 5 balls \\
		\hline
 	   2  & 8 & 16\\
 	   4  & 18 & 54\\
 	   6  & 32 & 128\\
 	   8  & 50 & 250\\
 	  10  & 72  & 432\\
 	  12  &  98 & 686\\
 	  14  &  128 & 1024\\
 	  16  &  162 & 1458\\
\hline
\end{tabular}
\caption{Cardinality of the ORTHOCUB support for the two integration examples.}
\label{cards}
\end{table}

\begin{table} 
\centering
\begin{tabular}{| c || c | c | c || c | c | c ||}
		\hline 
		ADE $n$ & spl bnd & 5 balls \\
		\hline
 	   2  & 1.22 & 1.57 \\
 	   4  & 1.15 & 1.43 \\
 	   6  & 1.07 & 1.28 \\
 	   8  & 1.08 & 1.27 \\
 	  10  &  1.07 & 1.21  \\
 	  12  &  1.07 & 1.18 \\
 	  14  &  1.07 & 1.19 \\
 	  16  &  1.06 & 1.16 \\
\hline
\end{tabular}
\caption{The stability ratio $\|\mathbf{w}\|_1/|\sum_j w_j|$ of ORTHOCUB for the two integration examples.}
\label{ratios}
\end{table}

\begin{table} 
\centering
\begin{tabular}{| c || c | c | c || c | c | c ||}
		\hline 
		ADE $n$ & spl bnd & 5 balls \\
		\hline
 	   2  & 5e-04 & 5e-03\\
 	   4  & 4e-04 & 1e-02\\
 	   6  & 4e-04 & 3e-02\\
 	   8  & 4e-04 & 5e-02\\
 	  10  & 4e-04  & 9e-02\\
 	  12  & 5e-04  & 1e-01\\
 	  14  & 5e-04  & 2e-01\\
 	  16  & 6e-04  & 3e-01\\
\hline
\end{tabular}
\caption{Construction times of the ORTHOCUB weights for the two integration examples (in seconds:  Matlab version on a PC with a 2.7 GHz Intel Core i5 CPU and 16 GB of RAM).}
\label{times}
\end{table}

\subsection{Differential cubature}

 The application of ORTHOCUB to differentiation functionals like (\ref{diff-fun}) allows to compute {\em partial derivatives of multivariate polynomials as weighted sums by global sampling}, with no need of 
system solving for the weighting construction. Computing derivatives by weighted sums of function values is clearly useful for numerical differentiation, but is also at the core of the so-called Differential Quadrature (DQ) approach to the numerical solution of PDEs 
(often named Differential Cubature (DC) in its multivariate version). 

Indeed, the DQ/DC method, in particular its polynomial version, since the original paper by Bellman's group in 1972 \cite{BKC72}, has been widely developed during the last fifty years, as an alternative approach to other popular techniques for numerical PDEs like Finite Elements and Finite Volumes, with a vast and multifaceted literature especially in engineering applications; for the only purpose of illustration, we may refer the reader, among others, to the monography \cite{W15} with the references therein. 

In view of (\ref{P&Q})-(\ref{weights2}), the weights 
of ORTHOCUB for the differential operator (\ref{diff-fun}) become 
{\color{black}
$$
\mathbf{w}(P)=\mathbf{w}(P;\partial ^\alpha)=diag(\mathbf{z})V _\psi
\{\partial^{\alpha_1}_{x_1}\,\dots 
\,\partial^{\alpha_d}_{x_d} \psi_j(\Lambda^{-1}(P-C))\}
$$
\begin{equation} \label{diffweights}
=\lambda_1^{-\alpha_1}\dots\lambda_d^{-\alpha_d} \, diag(\mathbf{z})V_\psi \mathbf{m}_0(\Lambda^{-1}(P-C))
\end{equation}
where the ``reference differential moments'' appear
\begin{equation} \label{m0}
\mathbf{m}_0(Q)=\mathbf{m}_0(Q;\partial^\alpha)=\{\partial^{\alpha_1}_{t_1}\,\dots 
\,\partial^{\alpha_d}_{t_d} \psi_j(Q)\}\;,\;\;Q\in [-1,1]^d\;.
\end{equation}
}

In the case of the product Chebyshev polynomials chosen in subsection 3.1, such differential moments can be  computed, using known formulas for the derivatives of the univariate polynomials. 
{\color{black} 
Indeed, for the univariate ultraspherical polynomials $P_s^{(a,a)}(t)$ we have
\begin{equation} \label{ultra}
\frac{d^m}{dt^m}P_s^{(a,a)}(t)=\frac{\Gamma(2a+s+m+1)}{2^m\Gamma(2a+s+1)}\,P_{s-m}^{(a+m,a+m)}(t)\;,
\end{equation}
cf. e.g. \cite{R60}. Then, 
for the product bivariate bases in subsection 3.1, using well-known identities for Jacobi polynomials, generalized binomial coefficients and the Gamma function (cf. e.g. \cite{R60,Salw18}),  we get after some calculations
\begin{itemize}

\item 2D Chebyshev of the first kind: $$a=-1/2\,,\,T_n(t)=\frac{P_n^{(-1/2,-1/2)}(t)}{P_n^{(-1/2,-1/2)}(1)}=\frac{P_n^{(-1/2,-1/2)}(t)}{{n-1/2\choose n}}$$
$$
\partial_{t_1}^{\alpha_1}\partial_{t_2}^{\alpha_2}\psi_j(Q)=\frac{\sqrt{(2-\delta_{h,o})(2-\delta_{k,o})}}{\pi}\,\frac{d^{\alpha_1}}{dt_1^{\alpha_1}}T_h(t_1)\,\frac{d^{\alpha_2}}{dt_2^{\alpha_2}}T_{k}(t_2)
$$
$$
=\frac{\sqrt{(2-\delta_{h,o})(2-\delta_{k,o})}}{2^{\alpha_1+\alpha_2}}\,\frac{\Gamma(h+1)}{\Gamma(h+1/2)}\,\frac{\Gamma(h+\alpha_1)}{\Gamma(h)}\,\frac{\Gamma(k+1)}{\Gamma(k+1/2)}\,\frac{\Gamma(k+\alpha_2)}{\Gamma(k)}
$$
\begin{equation}\label{2D-cheb29}
\times\,P_{h-\alpha_1}^{(\alpha_1-\frac{1}{2},\alpha_1-\frac{1}{2})}(t_1)\,P_{k-\alpha_2}^{(\alpha_2-\frac{1}{2},\alpha_2-\frac{1}{2})}(t_2)
\end{equation}

\item 3D Chebyshev of the first kind:
$$
\partial_{t_1}^{\alpha_1}\partial_{t_2}^{\alpha_2}\partial_{t_3}^{\alpha_3}\psi_j(Q)=\frac{\sqrt{(2-\delta_{h,o})(2-\delta_{k,o})(2-\delta_{\ell,o})}}{\pi^{3/2}}\,\frac{d^{\alpha_1}}{dt_1^{\alpha_1}}T_h(t_1)\,\frac{d^{\alpha_2}}{dt_2^{\alpha_2}}T_{k}(t_2)\,\frac{d^{\alpha_3}}{dt_3^{\alpha_3}}T_{k}(t_3)
$$
$$
=\frac{\sqrt{(2-\delta_{h,o})(2-\delta_{k,o})(2-\delta_{\ell,o})}}{2^{\alpha_1+\alpha_2+\alpha_3}}
\,\frac{\Gamma(h+1)}{\Gamma(h+1/2)}\,\frac{\Gamma(h+\alpha_1)}{\Gamma(h)}\,\frac{\Gamma(k+1)}{\Gamma(k+1/2)}\,\frac{\Gamma(k+\alpha_2)}{\Gamma(k)}
$$
\begin{equation} \label{3D-cheb}
\times \,\frac{\Gamma(\ell+1)}{\Gamma(\ell+1/2)}\,\frac{\Gamma(\ell+\alpha_3)}{\Gamma(\ell)}\,P_{h-\alpha_1}^{(\alpha_1-\frac{1}{2},\alpha_1-\frac{1}{2})}(t_1)\,\,P_{k-\alpha_2}^{(\alpha_2-\frac{1}{2},\alpha_2-\frac{1}{2})}(t_2)\,P_{\ell-\alpha_3}^{(\alpha_3-\frac{1}{2},\alpha_3-\frac{1}{2})}(t_3)
\end{equation}
\end{itemize}
both with the convention that $P_k^{(\cdot,\cdot)}$ is zero for $k<0$, and in addition $\Gamma(x)/\Gamma(0)=0$ for $x>0$ and $\Gamma(0)/\Gamma(0)=1$, so that the formulas are well-defined also when one or more of the degrees $h,k,\ell$ are zero, for any possible value of $\alpha_1,\alpha_2,\alpha_3\geq 0$.}
With the formulas above, in view of (\ref{diffweights})-(\ref{m0}) {\em by a single matrix-vector product} we can compute the action of any linear partial differential operator on polynomials.
\vskip0.5cm
\begin{Remark}
Computation of the differential moments above requires two main steps. One is the computation of univariate Jacobi polynomials, that can be done efficiently via their three-term recurrence relation. The other is the computation of ratios of Gamma functions. Ratios of the form $\Gamma(u+v)/\Gamma(u)$ with integer $u,v>0$ are simply the products $u(u+1)\dots(u+v-1)$, also known as Pochammer symbols $(u)_v$. On the other hand, ratios of the form $\Gamma(u+1)/\Gamma(u+1/2)$ for integer $u>0$, and more generally ratios of Gamma functions, starting from 
the classical papers by Frame \cite{F49} and by Tricomi and Erd\'{e}ly \cite{TE51} on their asymptotic approximation, has been object of a specific literature during more than seventy years. We may quote for example \cite{Mo10}, where such ratios are accurately  and efficiently approximated for relatively large $u$ by $2m$-th roots of special $m$-th degree polynomials in $u$, with $1\leq m\leq 6$. For small or moderate values of integer $u>0$, we can simply compute the ratio via a Pochammer symbol, as $\Gamma(u+1)/\Gamma(u+1/2)=(\Gamma(u+1)/\Gamma(u))(\Gamma(u)/\Gamma(u+1/2))=u/(1/2)_u$. 
\end{Remark}
\vskip0.5cm
\begin{Remark}
It is worth mentioning, however, that there is an alternative approach, based substantially on the construction of suitable {\em differentiation matrices}, a well-known concept and tool in polynomial methods for PDEs; cf., e.g., the monographies \cite{T00,W15}. Indeed, by (\ref{diffweights})-(\ref{m0}) let us define the first order differentiation weights
\begin{equation} \label{diff-weights}
\mathbf{w}(P\,;\partial_{x_k})
=\lambda_k^{-1}\,diag(\mathbf{z})V_\psi \{\partial_{t_k}\psi_j(\Lambda^{-1}(P-C))\}
\in \mathbb{R}^M\;,
\end{equation}
and the weighting matrices
\begin{equation} \label{DM}
\mathcal{D}_k=\mathcal{D}_{\partial_{x_k}}=(d_{ij})=(w_j(P_i\,;\partial_{x_k}))\in \mathbb{R}^{M\times M}\;,\;\;k=1,\dots,d\;.
\end{equation}
Moreover, we set for convenience $\mathbf{f}=\{f(P_j)\}\in \mathbb{R}^M$ (the vector of the function values at the cubature nodes). Then for every $f\in \mathbb{P}_n$, using the fact that $\mathbf{w}(P\,;\partial_{x_k})$ is itself a polynomial in $\mathbb{P}_n$, it is easily checked that 
\begin{equation} \label{diff-matr}
\partial_{x_k}^{\alpha_k}f(P)=\langle\mathbf{w}(P\,;\partial_{x_k}),\mathcal{D}_k^{\alpha_k-1}\mathbf{f}\,\rangle\;,\;\;\{\partial_{x_k}^{\alpha_k}f(P_i)\}=\mathcal{D}_k^{\alpha_k}\mathbf{f}\;,
\end{equation}
where $\langle\cdot,\cdot\rangle$ denotes the scalar product in $\mathbb{R}^M$. In such a way by a suitable sequence of matrix-vector products we can compute the action of any linear partial differential operator on polynomials. 
\end{Remark}
\vskip0.5cm
{\color{black} In the present implementation of differentiation by ORTHOCUB, we have privileged the ``special functions'' approach (\ref{ultra})-(\ref{3D-cheb}) with respect to the ``differentiation matrices'' approach (\ref{diff-weights})-(\ref{diff-matr}), essentially because it requires the storage and application of a single matrix. Nevertheless, the package could be complemented in the future with a not big effort, to include the construction of differentiation matrices and the corresponding implementation of differential cubature. 
Moreover, the present starting version is limited to the computation of first, second and mixed partial derivatives, that are sufficient to 
compute relevant differential operators like gradients, Jacobians, Hessians, as well as Laplacians and other second-order elliptic operators.}

\begin{figure}[h]
	\centering
		\includegraphics[scale=0.38,clip]{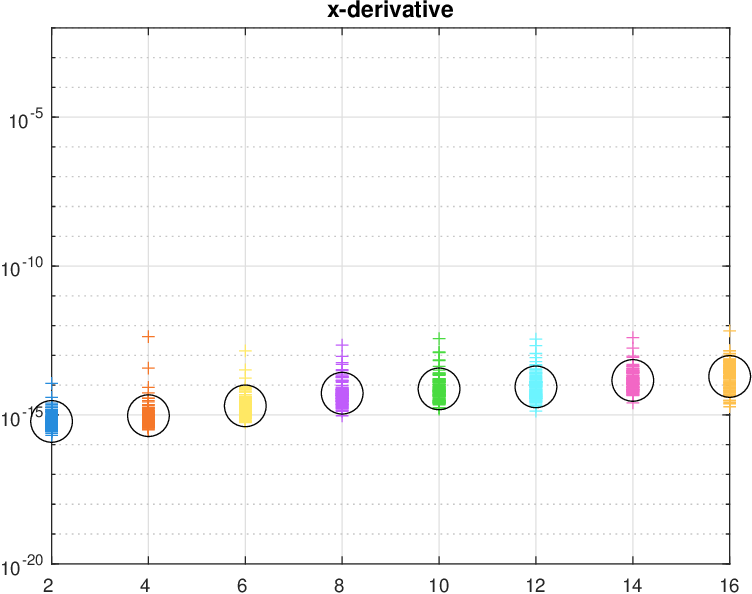}
		\includegraphics[scale=0.38,clip]{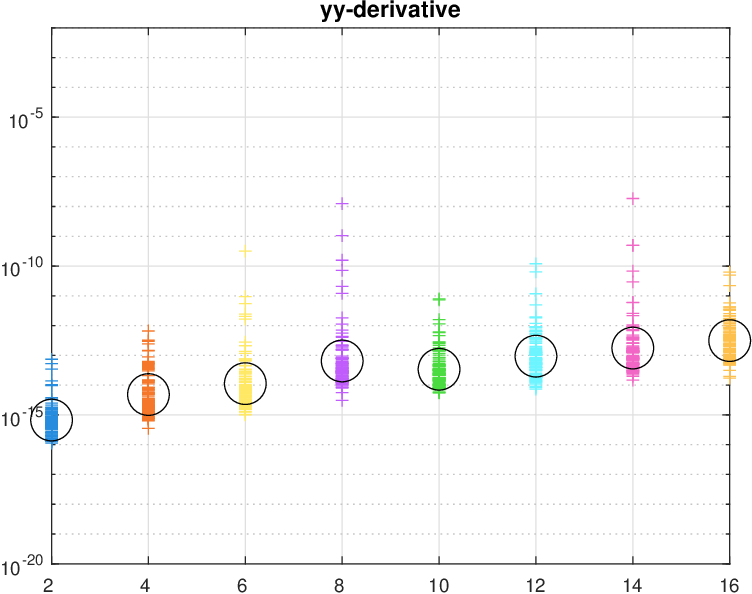}
        \includegraphics[scale=0.38,clip]{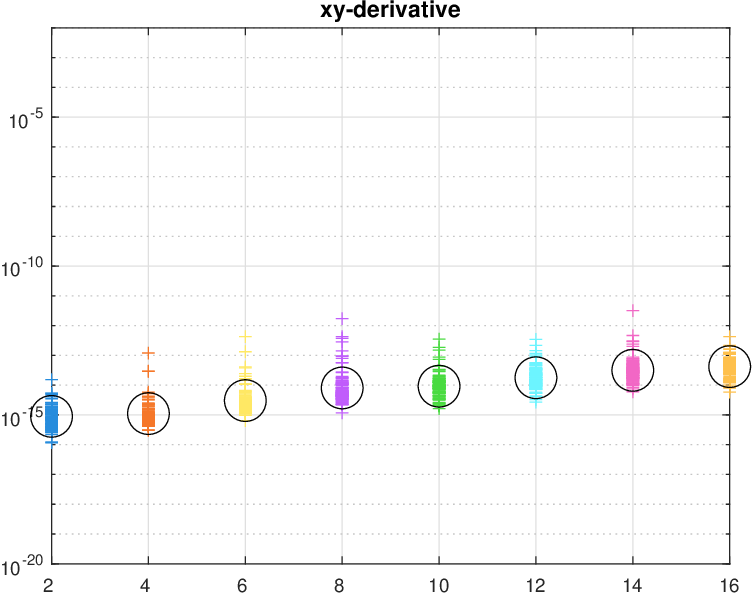}
	\caption{{\color{black}Small crosses: relative differentiation errors in the 2-norm on the first 100 Halton points in $[-1,1]^2$, for 100 trials of random polynomials $p_n(x,y)=(c_0+c_1 x+c_2 y)^n$, $n=2,4,\dots,16$. Circles: geometric mean of the relative errors.}}
	\label{partial2D}
\end{figure}

\begin{figure}[h]
	\centering
		\includegraphics[scale=0.38,clip]{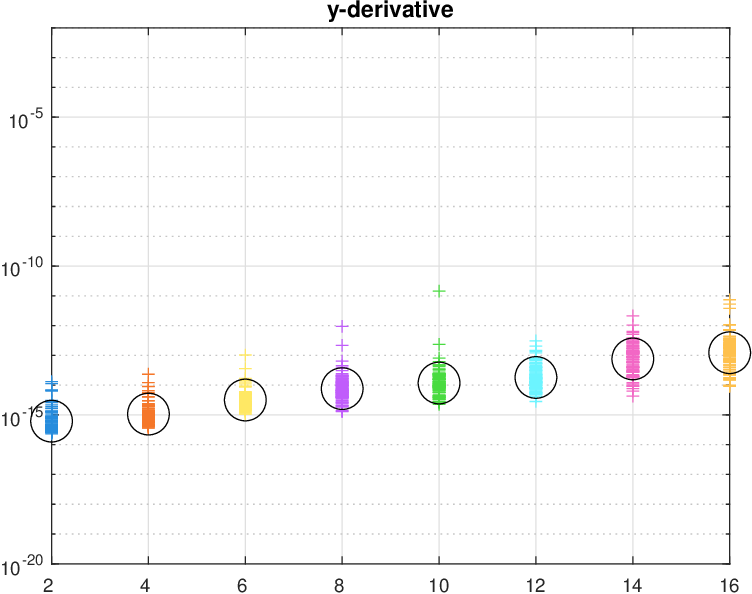}
		\includegraphics[scale=0.38,clip]{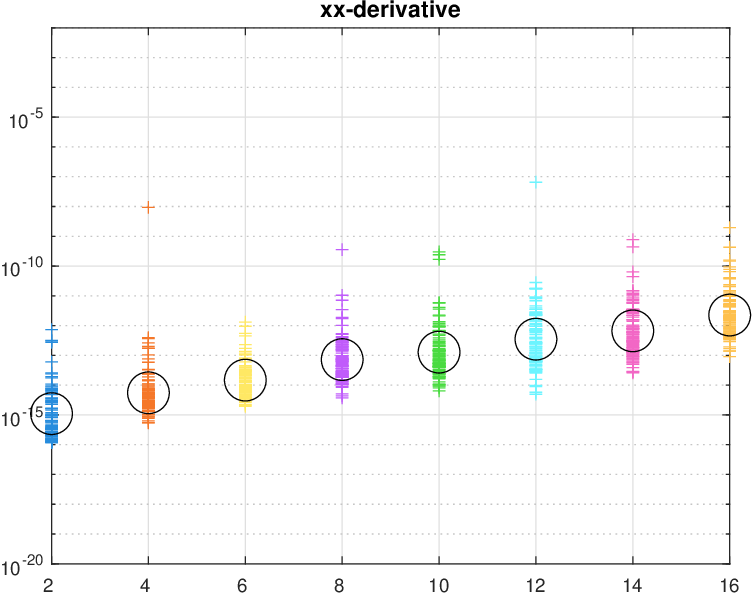}
        \includegraphics[scale=0.38,clip]{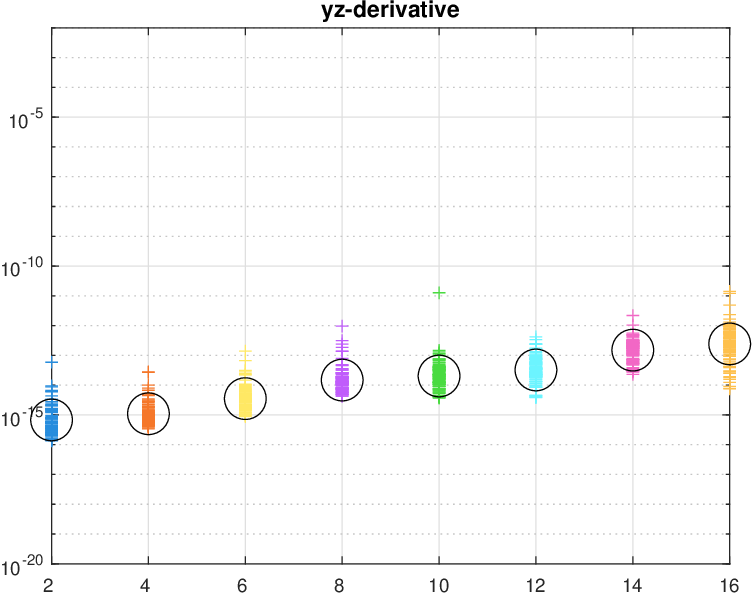}
	\caption{{\color{black}As in Figure \ref{partial2D} for the random polynomials $p_n(x,y,z)=(c_0+c_1 x+c_2 y+c_3z)^n$ in $[-1,1]^3$.}}
	\label{partial3D}
\end{figure}

In order to test the accuracy of ORTHOCUB for polynomial differentiation 
in 2D and 3D, we compute on the reference boxes $[-1,1]^d$, $d=2,3$, first and second partial derivatives of polynomials with uniform random coefficients in $(0,1)$. The results are collected in Figures \ref{partial2D}-\ref{partial3D} and show a geometric mean of the relative errors that departs up to about 4 orders of magnitude from machine precision increasing the degree, a classical phenomenon in numerical differentiation due to the intrinsic instability inherited from the differential operators. Such a degradation of the precision can be explained by looking at the ``Lebesgue constants'' of differential cubature, namely 
$$
\lambda_n(\partial^\alpha;B)=\max_{P\in B}{\|\mathbf{w}(P;\partial^\alpha)\|_1}=\lambda_1^{-\alpha_1}\dots\lambda_d^{-\alpha_d}\,\lambda_n(\partial^\alpha;[-1,1]^d)
$$
$$
=\lambda_1^{-\alpha_1}\dots\lambda_d^{-\alpha_d}\,\max_{Q\in [-1,1]^d}{\|\mathbf{w}(Q;\partial^\alpha)\|_1}
$$
\begin{equation} \label{lebdiff}
=\lambda_1^{-\alpha_1}\dots\lambda_d^{-\alpha_d}\,\max_{Q\in [-1,1]^d}{\|diag(\mathbf{z})V_\psi \mathbf{m}_0(Q)\|_1}\;,\;\;n\geq 0\;,
\end{equation}
(cf. (\ref{diffweights})-(\ref{m0})), that we have approximated heuristically taking the maximum on a large number of Halton points 
and reported in Figure \ref{lebdiff} for first, second and mixed partial derivatives. We recall that for $\alpha=(0,\dots,0)$ we get the Lebesgue constant of hyperinterpolation in a box, which is invariant under affine transformations and is known to be $\mathcal{O}(\log^d(n))$, cf. \S 3.2. We also notice that the Lebesgue constants of ORTHOCUB differentiation turn out to be the same for all first partial derivatives, for all second partial derivatives and for all mixed partial derivatives, a phenomenon due to symmetries of the basis and of the domain. For $\alpha_1+\dots+\alpha_d=1$ (first partial derivatives) the observed growth appears nearly quadratic, while for $\alpha_1+\dots+\alpha_d=2$ (second and mixed partial derivatives) nearly quartic in $n$, a fact that looks naturally  related to Markov inequality (cf. e.g. \cite{D92,Sh04}) applied (directionally) to the hyperinterpolation polynomial. 
We reserve to a future work a full error and stability analysis of ORTHOCUB beyond polynomial differentiation, that is for global numerical differentiation of regular functions. 

Finally, we report the CPU times for the construction of ORTHOCUB differential cubature weights for degrees $n=2,4,\dots,16$. The average time per point over a large number of trials in the computation of first, second and mixed partial derivatives on $10^4$ Halton points, is of the order of $10^{-6}-10^{-5}$ seconds in 2D, and $10^{-5}-10^{-4}$ seconds in 3D (Matlab version on a PC with a 2.7 GHz Intel Core i5 CPU and 16 GB of RAM).

\begin{figure}[h]
	\centering
		\includegraphics[scale=0.40,clip]{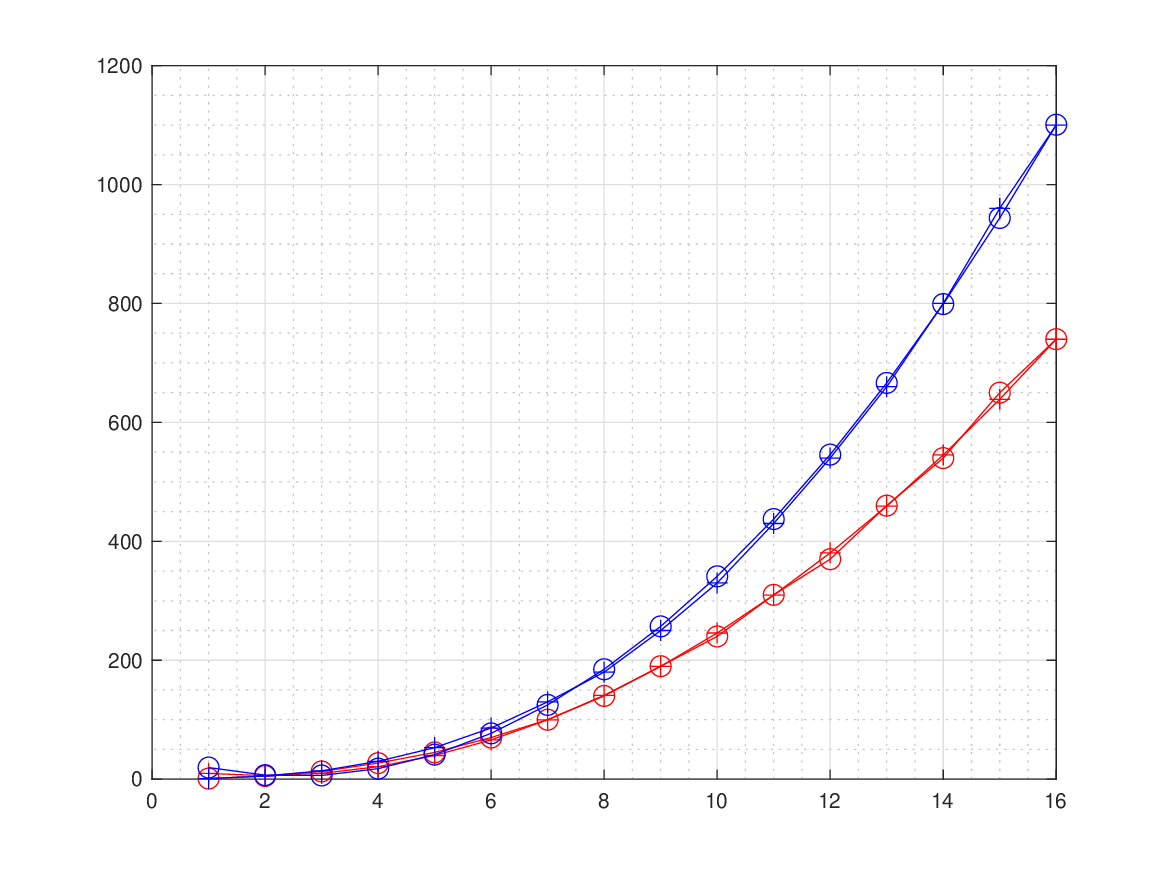}\\
        \includegraphics[scale=0.40,clip]{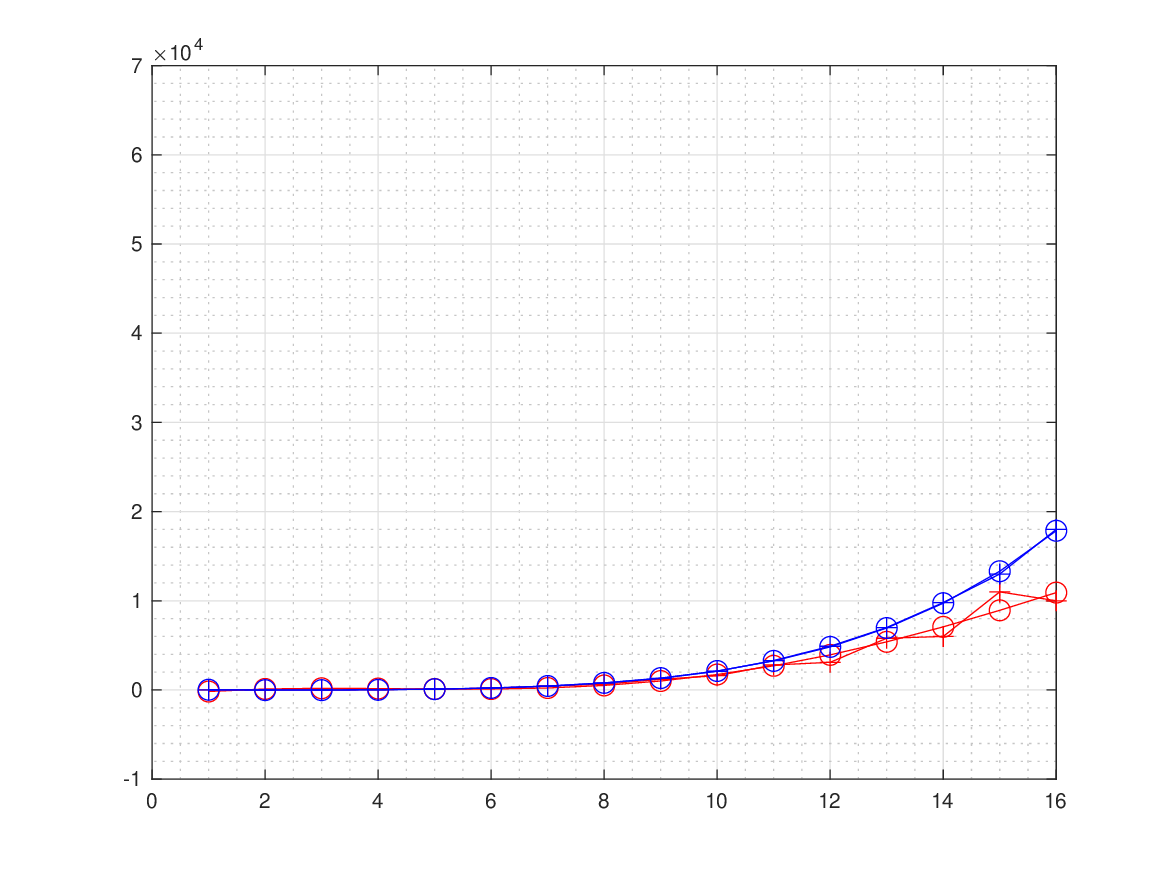}
		\includegraphics[scale=0.40,clip]{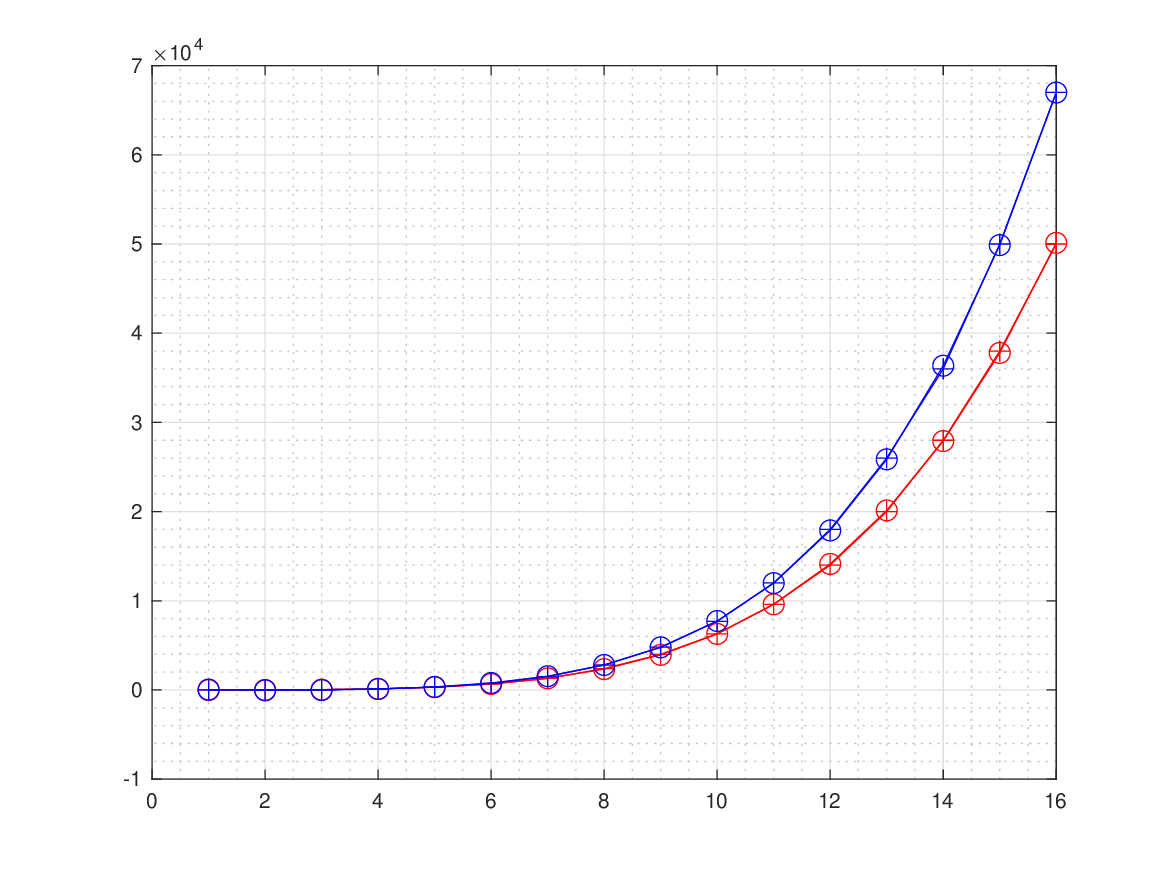}
        \caption{The ``Lebesgue constants'' (\ref{lebdiff}) of differential cubature 
        in $[-1,1]^2$ (red) and $[-1,1]^3$ (blue) with polynomial degree $n=2,4,\dots,16$: top: first partial derivatives (crosses), with their quadratic least-square fitting (circles); bottom: mixed (left) and second (right) partial derivatives (crosses), with their quartic least-square fitting (circles).}
        \label{lebdiff}
\end{figure}

\section{Software}

After the introduction to the main theoretical issues and discussion of numerical results, we describe the routines and demos used to perform these tests. They are implemented in Matlab and Python and available respectively at \cite{S25} and \cite{R25}. 

\subsection{Description of the main codes}
In what follows, we mention the routines that are of main importance for our numerical tests.  
\begin{itemize}
\item {\tt{cheap\_startup}}: given the algebraic degree of precision {\it{ade}}, it computes a low cardinality cubature rule {\it{rule\_ref}} on the reference square $[-1,1]^2$ or cube $[-1,1]^3$ with respect to the product Chebyshev weight (\ref{cheb2D}) or (\ref{cheb3D}). They correspond, respectively, to the Morrow-Patterson-Xu rules {\cite{CDMMV06}} and the formulas proposed in {\cite{DMVX09}}.

Next, calling the subroutine {\tt{dCHEBVAND\_orthn}}, it evaluates the orthonormal Chebyshev basis $\{\psi_j\}$  of total degree {\it{ade}} (see \S 3.1) at the nodes, storing the results in the Vandermonde matrix {\it{V\_ref}}. The sequence of indices $(h,k)$ or $(h,k,l)$ (see (\ref{cheb2D}) and (\ref{cheb3D})), defining the ordering of the basis elements, is listed in the matrix {\it{basis\_indices}}.

We stress that this routine is  independent of the functional $\mathcal{L}$ to be used and is only related to the ADE of the reference cubature rule.

\item {\tt{cheap\_rule}}: given the matrix {\it{dbox}} of dimension $2 \times d$ in which the $i$ column contains the extrema of the bounding box with respect to the $i$-th variable, the moments $m_j$ storing $\mathcal{L}(\psi_j(\Lambda^{-1}(\cdot-C)))$, as well as {\it{rule\_ref}}, {\it{V\_ref}} computed by 
{\tt{cheap\_startup}}, it determines the {\it{nodes}} $X$ and the {\it{weights}} {\bf{w}} of the ORTHOCUB cubature rule.

\item {\tt{dCHEBVAND\_orthn}} evaluates the Vandermonde matrix on the set $X \subset \Omega \subseteq B$, relatively to the scaled Chebyshev basis $\{\phi_j\}_j$ defined in (\ref{10}). As input, it is requires the variable {\it{dbox}} describing $B$ and the sequence of degree indices {\it{basis\_indices}}.

\item {\tt{vandermonde\_jacobi}} allows the evaluation of the Jacobi polynomials in (\ref{2D-cheb29}) and (\ref{3D-cheb}). This general purpose function takes into account a specific normalization of the basis, e.g. monic, orthonormal as well classical ones depending on the weight.

\item {\tt{cub\_square\_mpx}}, {\tt{cub\_cube\_mpx}} are low cardinality cubature rules, respectively on the unit-square $[-1,1]^2$ and on the cube $[-1,1]^3$ relatively to the product Chebyshev measure.
\item {\tt{mom\_derivative\_2D}}, {\tt{mom\_derivative\_3D}} produce the partial derivatives with respect to a variable specified by the user of the polynomial basis $\{\phi_j\}$ on a  specific 2D or 3D mesh. Depending on the mesh, a suitable bounding box $B$ must be defined via the variable {\it{dbox}}.  
\end{itemize}
As for the examples regarding numerical cubature, the algorithms require some additional functions.
\begin{itemize}
\item {\tt{compute\_spline\_boundary}} is useful in the case of the spline-curvilinear domains. It provides the description of its boundary, given as inputs the abscissae of the vertices $XV$ and the relative ordinates $YV$, a variable {\it{spline\_parms}} storing the order of the piecewise splines and the vertices involved. In case the spline is cubic, {\it{spline\_type}} fixes the additional conditions (e.g. natural, periodic). As output, the routine supplies the vector of structures {\it{Sx}} and {\it{Sy}} that determine the piecewise splines $x=x(t)$, $y=y(t)$ parametrizing the boundary.
\item {\tt{spline\_chebmom}}, given the polynomial degree {\it{n}}, the vectors of structures {\it{Sx}}, {\it{Sy}}, the basis ordering {\it{basis\_indices}}, computes the moments of the basis $\{\psi_j\}$ and, if required, of $\{\phi_j\}$. This purpose is obtained by applying Green theorem, that is the numerical integration of a certain piecewise polynomial function on the boundary.
\item {\tt{QMC\_union\_balls}}, given the vectors of {\it{centers}} and {\it{radii}} defining the balls $\mathcal{B}_i$, $i=1,2,\ldots$, computes {\it{card}} Halton points $X_B$ in the bounding box $B$ of the integration domain $\Omega=\cup_{i} \mathcal{B}_i$. Next, after the application of an in-domain function, provides as output the nodes and the weights of a QMC rule in $\Omega$.
\end{itemize}

\subsection{Description of the demos}
In order to make the examples replicable, we have implemented the following demos, used for the numerical experiments above.
\begin{itemize}
\item {\tt{demo\_cubature\_splines}} illustrates the basic usage of the ORTHOCUB technique to approximate definite integrals on a spline-curvilinear domain $\Omega_1$ (see Figure \ref{spline}).

\item {\tt{demo\_cubature\_ade\_splines}}, fixed a degree of exactness $n=2,4,\dots,16$, approximates via the rules proposed in this work 100 integrals of random polynomials of the form
$$
p_n(x,y)=(c_0+c_1 x+c_2y)^n.
$$
Each exact integral $\int_{\Omega_1} p_n(x,y) dx dy$ is computed via Green theorem.
Finally it plots the single relative errors and the value of their geometric mean.

\item {\tt{demo\_cubature\_sumweights\_spline}} first computes ORTHOCUB rules with degree of exactness $n=2,4,\dots,16$ for integration on $\Omega_1$ and then their stability ratios as in Table 2.

\item {\tt{demo\_cubature\_weights\_spline}}, first computes ORTHOCUB rules with degree of exactness $n=2,4,\dots,16$ for integration on $\Omega_1$, and then sorts the weights, illustrating their distribution as in Figure \ref{spline}.

\item {\tt{demo\_cubature\_QMC}} shows the basic usage of the ORTHOCUB technique to approximate definite integrals  on the domain $\Omega_2$ that is union of 5 balls (see Figure \ref{balls}).                       

\item {\tt{demo\_cubature\_ade\_QMC}} determines via {\tt{QMC\_union\_balls}} two QMC rules $S_L$, $S_H$ on $\Omega_2$, based respectively on $10^5$ and $10^7$ Halton points in the bounding box $B$.
Next, fixed a degree of precision $n=2,4,\dots,16$, approximates via the rules proposed in this work 100 integrals of random polynomials of the form
$$
p_n(x,y,z)=(c_0+c_1 x+c_2 y +c_3 z)^n.
$$
In particular, takes into account $S_H(p_n)$ as reference value, while the moment vector is $\{S_L({\phi_i})\}$, $1\leq i\leq N$.
Finally it plots the single relative errors and the value of their geometric mean.

\item {\tt{demo\_cubature\_sumweights\_QMC}} computes  ORTHOCUB rules with degree of exactness $n=2,4,\dots,16$ for QMC integration on $\Omega_1$ and their stability ratios as in Table 2.                   
               
\item {\tt{demo\_cubature\_weights\_QMC}} computes ORTHOCUB rules with degree of exactness $n=2,4,\dots,16$ for QMC integration on $\Omega_2$, sorts the weights and illustrates their distribution as in Figure \ref{balls}.                                                      
\item {\tt{demo\_derivative\_ade\_2D}} computes by the ORTHOCUB technique described above the first and second order derivatives of random polynomials of the form
$$
p_n(x,y)=(c_0+c_1 x+c_2 y)^n
$$
on a mesh of the first $10000$ Halton points $\{P_k\}$ in $[-1,1]^2$, for $n=2,4,\dots,16$. For any exponent $n$ and point of mesh $P_k$, we determine by an ORTHOCUB rule on $[-1,1]^2$ defined by {\tt{mom\_derivative\_2D}}  the numerical approximation of these derivatives on $P_k$. Since the exact value is known analytically, we can easily establish the relative error in norm 2. 

\item {\tt{demo\_derivative\_ade\_3D}} computes by the ORTHOCUB technique described above the first and second order derivatives of random polynomials of the form
$$
p_n(x,y,z)=(c_0+c_1 x+c_2 y +c_3 z)^n
$$
on a mesh of the first $10000$ Halton points $\{P_k\}$ in $[-1,1]^3$, for $n=2,4,\dots,16$. For any exponent $n$ and point of mesh $P_k$, we determine by an ORTHOCUB rule on $[-1,1]^3$ defined by {\tt{mom\_derivative\_3D}} the numerical approximation of these derivatives on $P_k$. Since the exact value is known analytically, we can easily establish the relative error in norm 2.

\item {\tt{demo\_derivative\_weights\_2D}} after defining a mesh of the first $10000$ Halton points $\{P_k\}$ in $[-1,1]^2$, it computes for each of them the weights of a cubature rule in $[-1,1]^2$ with a certain ADE, relative to the first or second order partial derivative; next it calculates for each derivative the maximum sum of the absolute value of those weights, finally plotting the results, i.e. the ``Lebesgue constants'' of differential cubature in $[-1,1]^2$.

\item {\tt{demo\_derivative\_weights\_3D}} after defining a mesh of the first $10000$ Halton points $\{P_k\}$ in $[-1,1]^3$, it computes for each of them the weights of a cubature rule in $[-1,1]^3$ with a certain ADE, relative to the first or second order partial derivative; next it calculates for each derivative the maximum sum of the absolute value of those weights, finally plotting the results, i.e. the ``Lebesgue constants'' of differential cubature in $[-1,1]^3$.

\end{itemize}

\vskip0.5cm 
\noindent
{\bf Acknowledgements.} 

Work partially supported by the DOR funds of the University of Padova and by the INdAM-GNCS. 
This research has been accomplished within the Community of Practice 
``Green Computing" of the Arqus European University Alliance, the RITA ``Research ITalian network on Approximation", and the SIMAI Activity Group ANA\&A.

\end{document}